\documentclass[aps,prl,reprint,twocolumn,superscriptaddress,showpacs,floatfix,nofootinbib]{revtex4-1}
\usepackage{amsmath,amsfonts,graphicx,color,yfonts,hyperref,bbm}
\def\bra{\langle}
\def\ket{\rangle}
\newcommand{\trento}{T$\mathrel{\protect\raisebox{-2.1pt}{R}}$ENTo}

\begin{document}

\title{Observing the deformation of nuclei with relativistic nuclear collisions}

\author{Giuliano Giacalone}
\affiliation{Universit\'e Paris-Saclay, CNRS, CEA, Institut de physique th\'eorique, 91191, Gif-sur-Yvette, France.}

\begin{abstract}
I show that particle collider experiments on relativistic nuclear collisions can serve as direct probes of the deformation of the colliding nuclear species. I argue that collision events presenting very large multiplicities of particles and very small values of the average transverse momentum of the emitted hadrons probe collision geometries in which the nuclear ellipsoids fully overlap along their longer side. By looking at these events one selects interaction regions whose elliptic anisotropy is determined by the deformed nuclear shape, which becomes accessible experimentally through the measurement of the elliptic flow of outgoing hadrons.
\end{abstract}

\maketitle

\section{introduction}

The majority of atomic nuclei are not spherical in their ground states, but present a quadrupole deformation. This property of nuclei can not be revealed directly in experiments~\cite{Raman:1201zz}, and our knowledge of it comes mostly from the results of theoretical calculations~\cite{Soma:2018qay} which are rarely confronted with experimental data. In this article, I propose to use nucleus-nucleus collision experiments at relativistic energies as a probe of nuclear structure.

Relativistic nuclear collisions are performed in the world's largest particle accelerator facilities, the BNL Relativistic Heavy Ion Collider (RHIC) and at the CERN Large Hadron Collider (LHC). By smashing two nuclei against each other at very high energy, one produces the quark-gluon plasma (QGP)~\cite{Busza:2018rrf}, the high-temperature state of strong-interaction matter. The QGP created in a nuclear collision exists for a very short time ($\approx 10^{-22}$~s) before transforming into thousands of particles that are observed in the detectors. 

\begin{figure}[t]
    \centering
    \includegraphics[width=.9\linewidth]{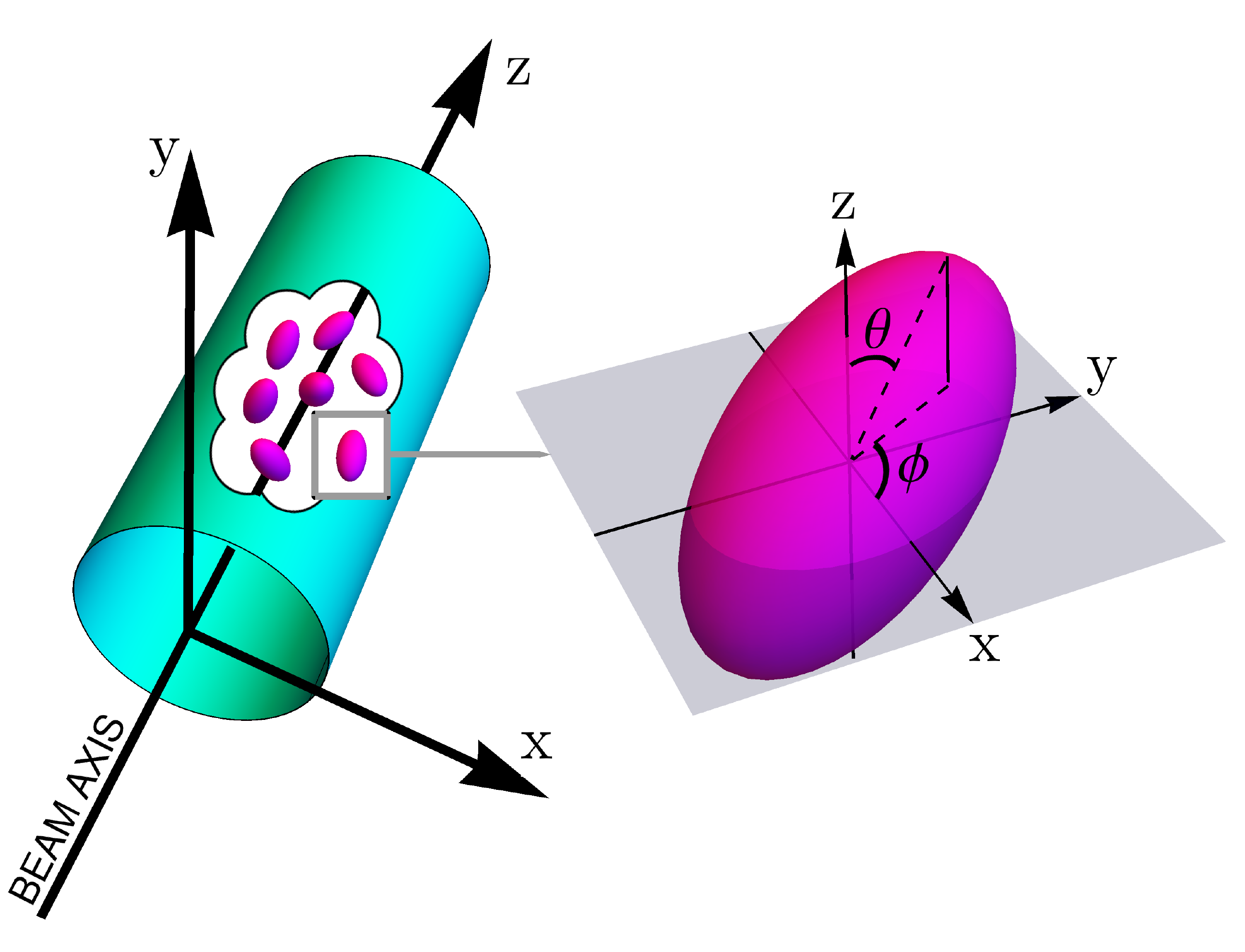}
    \caption{A bunch of deformed nuclei accelerated in a beam pipe. The beam axis runs along the $z$ axis. Individually, each nucleus has a random orientation in space, determined by a polar angle, $\theta$, and an azimuthal angle, $\phi$.}
    \label{fig:1}
\end{figure}

Remarkably enough, the distribution of these particles in momentum space and their mutual correlations~\cite{Luzum:2011mm} carry information about the geometric shape of the colliding nuclei. The reason is that the QGP is a hydrodynamic medium~\cite{Romatschke:2017ejr}, whose dynamics is governed by pressure-gradient forces:
\begin{equation}
\label{eq:euler}
    F = - \nabla P,
\end{equation}
where $F$ is the force per unit volume, and $\nabla P$ is the pressure gradient. The QGP is created at rest, and set in motion by these pressure gradients, which are determined by the geometry of the system. This geometry is in turn determined by how two nuclei overlap at the time of the interaction, a feature which depends on the spatial orientations of the nuclear axes (see Fig.~\ref{fig:1}).

These orientations are random, and generate nontrivial geometries of overlap, that leave distinct signatures in the distributions of final-state particles.
It has already been argued that several results obtained in collisions between deformed nuclei ($^{238}$U+$^{238}$U and $^{129}$Xe+$^{129}$Xe collisions) can only be explained by taking into account their quadrupole deformation~\cite{Adamczyk:2015obl,Sirunyan:2019wqp,Acharya:2018ihu,Aad:2019xmh}. However, these results were obtained by averaging the data sets over all possible orientations of the nuclear axes, thus reducing the sensitivity of the observables to the nuclear structure.

Here I present a procedure for analyzing nucleus-nucleus data that allows to look at \textit{frozen} collision geometries where both nuclear axes are perpendicular to the collision axis. In these configurations the shape of the QGP closely follows the shape of the colliding bodies, and the angular distribution of emitted particles becomes a sensitive probe of the deformed nuclear shapes.

\section{freezing nuclear orientations}

The density of matter in an axially-symmetric nucleus with a static
quadrupole deformation can be written in the form of a two-parameter Fermi distribution:
\begin{equation}
\label{eq:density}
    \rho({\bf x}',z') = \frac{\rho_0}{1+\exp \biggl\{ \frac{1}{a} \biggl[  \sqrt{|\textbf{x}'|^2+z'^2} - R \bigl(1 + \beta Y_{20} \bigr)  \biggr] \biggr\} },
\end{equation}
where $z'$ is the axis of the nucleus, $\textbf{x}'$ is a coordinate in the plane orthogonal to $z'$, and spherical symmetry is broken by the spherical harmonic that carries a dependence on the angle $\Theta$ between ${\bf x}'$ and $z'$, $Y_{20}=\sqrt[]{\frac{5}{16 \pi}}\bigl (3\cos^2 \Theta-1 \bigr)$. $a$ is the diffusiveness of the nucleus, $R$ is its average radius, and $\rho_0$ is the density of nuclear matter. The quadrupole deformation of the nuclear ellipsoid
is controlled by the parameter $\beta$. A nucleus is spherical
for $\beta=0$, prolate for $\beta>0$, and oblate for $\beta<0$.

Now, inject such a nucleus in the beam pipe of a particle collider. The laboratory frame is defined by the beam axis, $z$, and the plane orthogonal to it, ${\bf x}$, the so-called \textit{transverse} plane.  As illustrated in Fig.~\ref{fig:1}, each nucleus in the beam pipe is randomly oriented in space, so that the intrinsic frame of the nucleus and the laboratory frame differ in general by polar tilt, $\theta$, and by an azimuthal spin, $\phi$. The geometry of the collision of two nuclei, say A and B, in the laboratory frame is therefore determined by two polar tilts, $\theta_A$ and $\theta_B$, and two azimuthal spins, $\phi_A$ and $\phi_B$~\cite{Schenke:2014tga,Goldschmidt:2015kpa}. 

I describe now a method for analyses of nucleus-nucleus data that permits to isolate configurations where $\theta_A = \theta_B = \pi/2$ and $\phi_A=\phi_B$. I exploit the hydrodynamic nature of the QGP, recalling that the medium created in high-energy collisions is approximately invariant under longitudinal boosts~\cite{Bjorken:1982qr}, so that one can solve its dynamics only in the transverse plane, in the longitudinal slice at $z=0$, where 0 is the interaction point of two nuclei.

The energy per particle in a thermodynamic medium is proportional to the temperature of the system. In the ultrarelativistic limit, the energy of a particle coincides with its momentum. Therefore, when a QGP decouples to particles one naturally expects that the transverse momentum, $p_t=\sqrt{p_x^2 + p_y^2}$, carried by these particles is proportional to the temperature of the medium at freeze out. Suppose now to have two QGPs presenting the same total entropy but different volumes. The one which is smaller in size is denser, has larger temperature, and therefore decouples to particles that carry more transverse momentum to the final state~\cite{Broniowski:2009fm,Gardim:2019xjs}. This simple argument is confirmed by hydrodynamic simulations~\cite{Mazeliauskas:2015efa,Bozek:2017elk}, where, at fixed total entropy, the initial size of the system, $R$, and the average transverse momentum of the produced hadrons, $\bar p_t$, are strongly anti-correlated.

\begin{figure}[t]
    \centering
    \includegraphics[width=.9\linewidth]{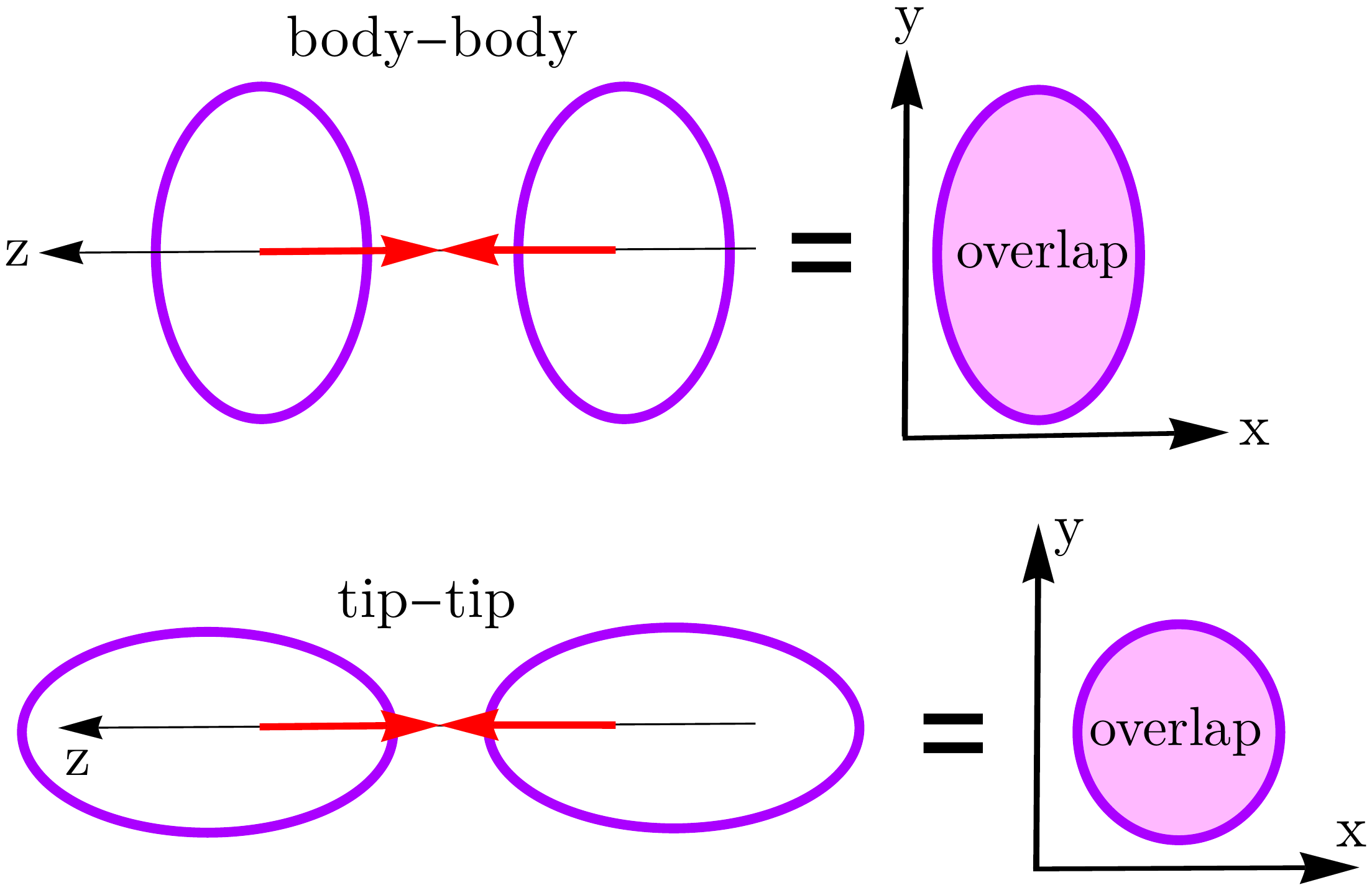}
    \caption{Tip-tip and body-body configurations of non-spherical nuclei colliding at zero impact parameter. On the right I show resulting transverse areas of overlap at fixed $z$.}
    \label{fig:2}
\end{figure}
Now, let us have a look at Fig.~\ref{fig:2}. Body-body collisions present $\theta_A=\theta_B=\frac{\pi}{2}$, while and tip-tip collisions present $\theta_A=\theta_B=0$. The area of overlap is larger in body-body collisions. 
If one takes $R$ as the radius of the system weighted by the entropy density, $s({\bf x})$,
\begin{equation}
\label{eq:rad}
    R^2 = \frac{\int_{\bf x} |{\bf x}|^2 s({\bf x}) }{\int_{\bf x} s({\bf x}) }
\end{equation}
then, at fixed total entropy, body-body collisions present the larger values of $R$, and therefore smaller values of $\bar p_t$ than tip-tip collisions. This result provides an experimental handle on the orientation of the colliding nuclei. First, choose events at fixed total entropy, which can be done experimentally by selecting events that present the same number of emitted particles, or \textit{multiplicity}. Then, sort these events according to the $\bar p_t$ of the produced hadrons. Events with abnormally low values of $\bar p_t$ correspond to fully overlapping body-body collisions.

I perform an explicit application of this selection procedure in simulations of the collision process. I use the phenomenologically-successful \trento{} model of initial conditions~\cite{Moreland:2014oya}. In this model, the profile of entropy density created in the interaction of nuclei A and B behaves like $s({\bf x}) \propto \sqrt{T_A({\bf x}+{\bf b}/2)T_B({\bf x}-{\bf b}/2)}$, where $T_{A/B}({\bf x})$ is a Lorentz-boosted matter density, given by the integral of Eq.~(\ref{eq:density}) along the beam axis, $z$, and ${\bf b}$ is the impact parameter of the collision. \trento{} includes the effect of fluctuations, both at the level of the positions of the colliding nucleons, as well as in the amount of entropy that they produce. The entropy  produced  by  a  participant nucleon  is  distributed  according  to  a gamma distribution, which is tuned to data following the comprehensive phenomenological applications of Refs.~\cite{Giacalone:2017dud,Giacalone:2018apa}.

\trento{} provides, in each event, the value of $R$. To express my results as function of quantities that are measurable, I use the effective hydrodynamic framework of Refs.~\cite{Gardim:2019xjs,Gardim:2019brr} to convert the relative variation of $R$ into an approximation of the relative variation of $\bar p_t$ as follows:
\begin{equation}
\label{eq:relative}
    \frac{\delta \bar p_t}{\bra \bar p_t \ket} = -3c_s^2  \frac{\delta R}{\bra R \ket}  ,
\end{equation}
where $\delta \bar p_t = \bar p_t - \bra \bar p_t \ket$, $\delta R = R - \bra R \ket$, and angular brackets denote statistical averages in the multiplicity class. I recall that $\bar p_t$ is the average transverse momentum in \textit{one} event.
The speed of sound of the quark-gluon plasma, $c_s^2$, appears in the equation as it tells us how an increase in temperature translates into an increase in pressure in the medium. Following Ref.~\cite{Gardim:2019xjs}, I shall use $c_s^2\approx0.25$ at LHC energies, and that $c_s^2 \approx 0.19$ at RHIC energies.

I simulate $2\times10^7$ $^{238}$U+$^{238}$U collisions (see Tab.~\ref{tab:1}).
I restrict my analysis to collisions occurring at very small impact parameter (central collisions), where the geometry of the system is mostly determined by the orientation of the colliding bodies. Experimentally, the impact parameter is not an observable quantity, and the centrality of a collision is quantified by the amount of produced particles. Central collisions correspond to events displaying very large multiplicities in the final state. In the \trento{} model, this amounts to select events that present a very large initial total entropy. Therefore, I focus on a narrow class of events in the high-entropy tail of my sample, specifically, the 0-0.5\% most central events.
\begin{figure}[t]
    \centering
    \includegraphics[width=.75\linewidth]{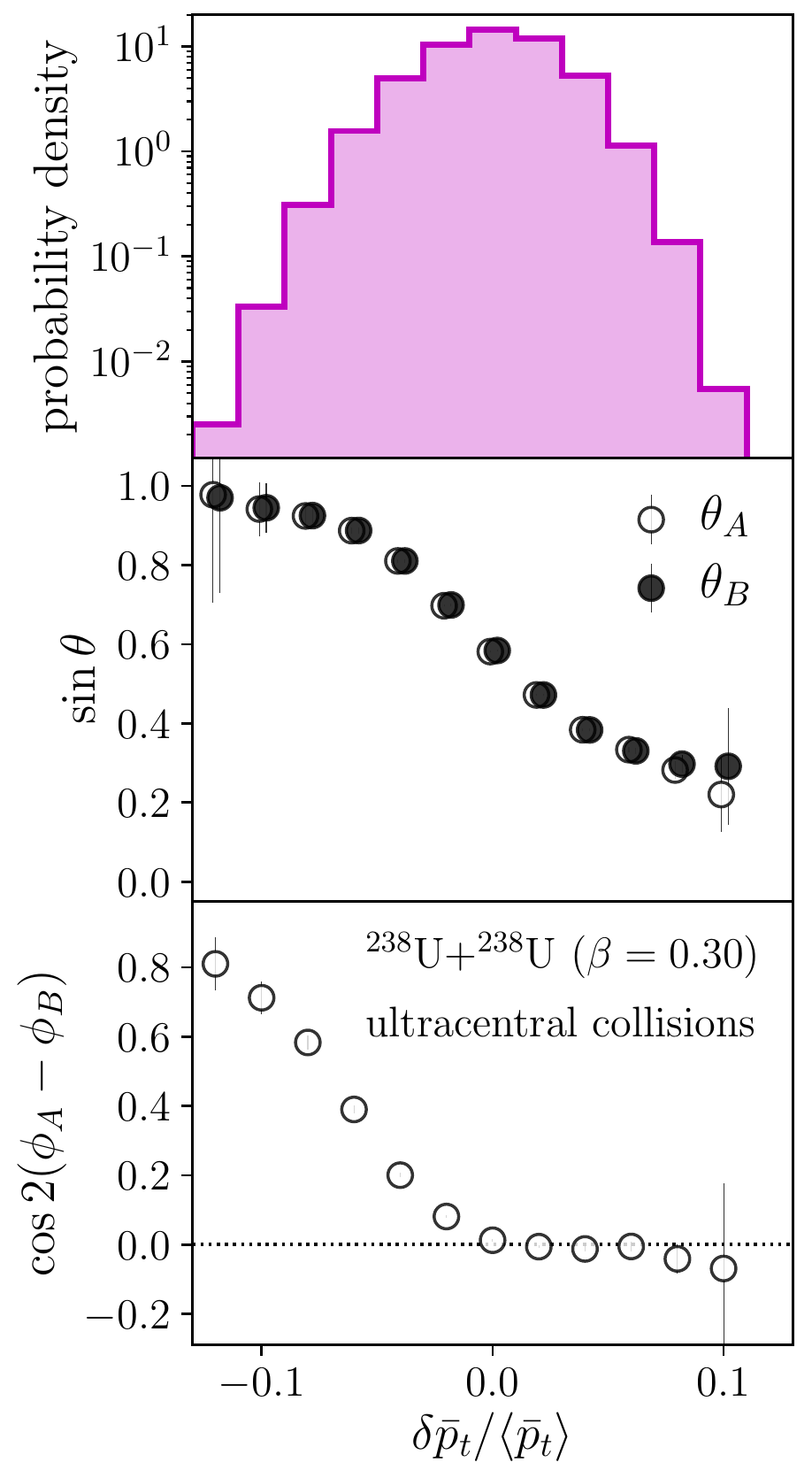}
    \caption{Effect of sorting ultracentral U+U collisions according to their $\bar p_t$. In the uppermost panel, I show the distribution of $\bar p_t$ around its average. In each bin of the histogram, I evaluate the average of the sine of the polar angles of the colliding nuclei (middle), and the average alignment between the azimuthal angles of the nuclei (lowermost panel).}
    \label{fig:3}
\end{figure}

I first show the dispersion of $\bar p_t$ around its average value in the upper panel of Fig.~\ref{fig:3}. We note that the distribution is skewed, with larger values of probability in the low-$\bar p_t$ tail. As it will be made clear below, this is a consequence of the positive value of $\beta$.

Moving on to the middle panel of Fig.~\ref{fig:3}, for each bin of the previous histogram I plot the values of $\sin \theta$ as function of $\bar p_t$ for both nuclei. The sinus grows very close to unity at low $\bar p_t$. This indicates that the events belonging to the low-$\bar p_t$ tail present $\theta_A=\theta_B=\pi/2$, confirming our expectations. Note that at large $\bar p_t$ the sine does not flatten around 0, which would correspond to the limit of tip-tip events, $\theta_A=\theta_B=0$. Closer investigation reveals that, in fact, the large-$\bar p_t$ tail selects events with large impact parameter, rather than tip-tip configurations.

Finally, in the lower panel of Fig.~\ref{fig:3} I compute the alignment of the two nuclei in the azimuthal plane by evaluating $\cos 2(\phi_A - \phi_B)$. At average and large transverse momentum, the correlator is consistent with zero. Remarkably enough the correlator approaches unity as one moves to lower $\bar p_t$ values, which implies almost perfect alignment between azimuthal angles, i.e., $\phi_A=\phi_B$.

In summary, the low-$\bar p_t$ tail of ultracentral events selects fully overlapping body-body collisions. I now show how the nuclear deformation can be observed in this selection of events.

\section{revealing nuclear deformation}
\begin{figure*}
    \centering
    \includegraphics[width=.65\linewidth]{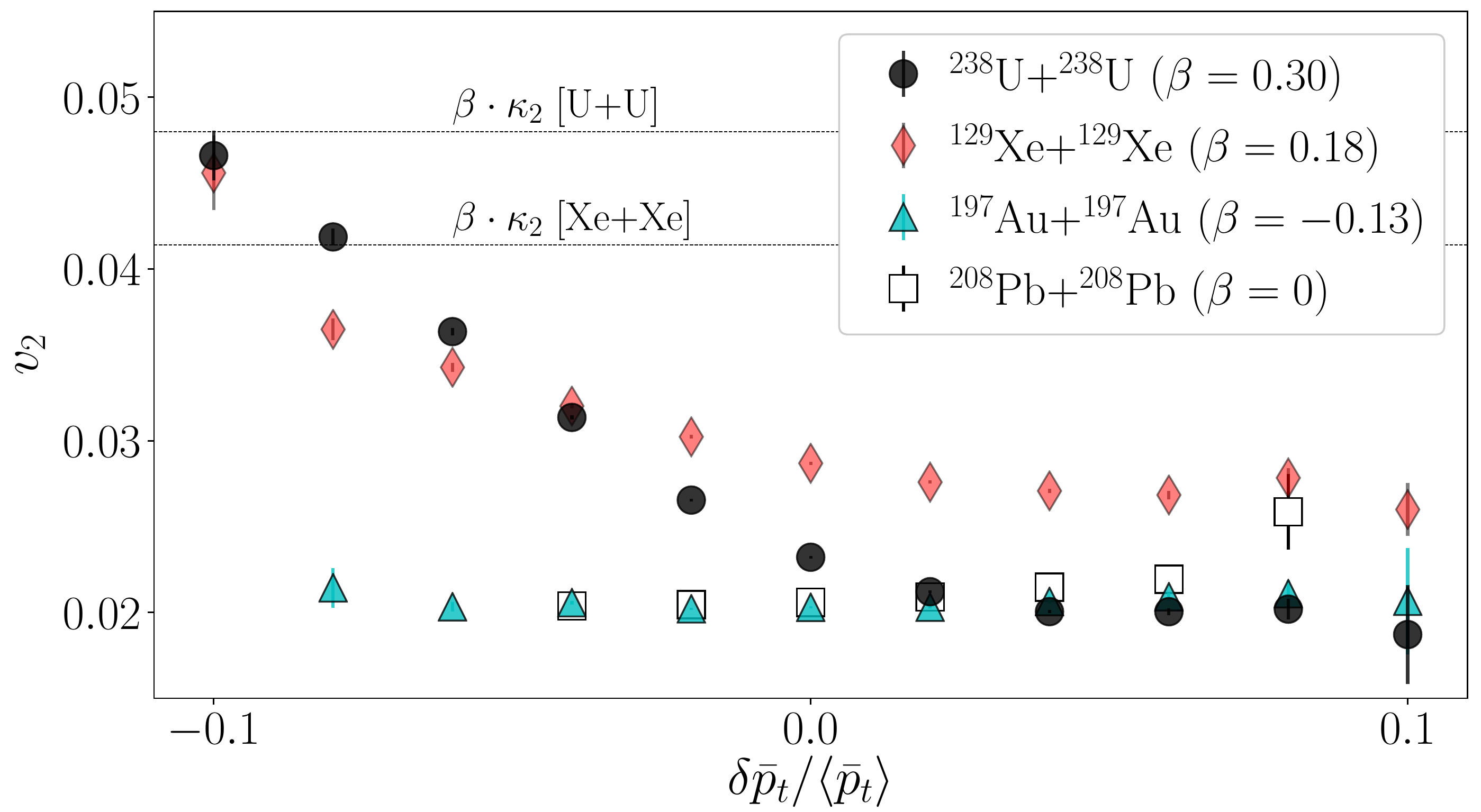}
    \caption{Elliptic flow as function of the average transverse momentum in ultracentral collisions of  $^{238}$U, $^{208}$Pb, $^{197}$Au, $^{129}$Xe nuclei. In each plot, by moving from right to left one gradually freezes the orientation of the colliding bodies towards the limit of body-body collisions. The horizontal dashed lines indicate the values of $v_2=\beta \kappa_2$ (i.e. $\varepsilon_2=\beta$).}
    \label{fig:4}
\end{figure*}
If the region of nuclear overlap has an elliptic shape like in the upper panel of Fig.~\ref{fig:2}, pressure gradients are larger along $x$ than along $y$, due to the smaller transverse size~\cite{Ollitrault:1992bk}. Equation~(\ref{eq:euler}) predicts, then, an asymmetry in the forces~\cite{Ollitrault:2008zz}:
\begin{equation}
\label{eq:dim3}
F_x > F_y,
\end{equation}
so that more momentum is built along $x$ than along $y$. This phenomenon is called \textit{elliptic flow}. Experimentally, it manifests as an angular imbalance in the momentum carried by the produced particles, corresponding to a $\cos (2\phi)$ modulation of the azimuthal particle spectrum:
\begin{equation}
    \frac{dN}{d\phi} \propto 1 + 2 v_2 \cos(2\phi),
\end{equation}
where $v_2$ quantifies the magnitude of elliptic flow.

Let us go, then, back to Fig.~\ref{fig:2}. While tip-tip collisions produce a quark-gluon plasma with a circular background geometry, body-body collisions have a manifest elliptical asymmetry, due to the deformation of the colliding bodies.
Equation ~(\ref{eq:dim3}) implies, then, that the largest elliptic flow is achieved in fully-overlapping body-body collisions.
Combining this argument with the previous observation that fully-overlapping body-body collisions are those presenting abnormally small values of $\bar p_t$, I conclude that nuclear deformation yields an enhancement of elliptic flow in the low-$\bar p_t$ tail of ultracentral events. 

I verify the validity of this picture in the \trento{} model. I shall not, though, compute elliptic flow by means of full hydrodynamic simulations. I use the fact that elliptic flow is a response to the initial \textit{eccentricity} of the medium~\cite{Teaney:2010vd}:
\begin{equation}
    \varepsilon_2 = \frac{|\int_{\bf x} {\bf x}^2  s({\bf x})|}{\int_{\bf x} |{\bf x}|^2 s({\bf x})}
\end{equation} 
where ${\bf x}^2$ in the numerator is here in complex notation, ${\bf x}^2=(x+iy)^2$. Hydrodynamic simulations~\cite{Niemi:2015qia} show that the following relation holds to a very good approximation in central nucleus-nucleus collisions:
\begin{equation}
    v_2 = \kappa_2 \varepsilon_2.
\end{equation}
$\kappa_2$ is a response coefficient that depends on the properties of the medium, such as its viscosity. Its value has been determined at both RHIC and LHC energies.

Within the same batch of events used in Fig.~\ref{fig:3}, I calculate $\varepsilon_2$ in U+U collisions as function of $\bar p_t$, and I rescale it by a factor $\kappa_2=0.16$~\cite{Giacalone:2018apa} to obtain the final-state elliptic flow. The result is shown as full circles in Fig.~\ref{fig:4}(a). The intuitive prediction is confirmed. Elliptic flow is enhanced in the low-$\bar p_t$ region, towards the limit of fully-overlapping body-body collisions. Note that $v_2$ at low $\bar p_t$ reaches a value close to $\beta\kappa_2$, which implies $\varepsilon_2\approx\beta$. This is not a coincidence. The rms eccentricity can in general be decomposed as $\sqrt{\bra \varepsilon_2^2 \ket} = \sqrt{\varepsilon_g^2 + \sigma^2}$, where $\sigma$ is the $\bar p_t$-independent eccentricity fluctuation, while $\varepsilon_g$ is the contribution from the intrinsic elliptic geometry of the system, which in these results originates entirely from the deformation of the colliding bodies. In body-body events, one has in fact $\varepsilon_g \approx \beta$, which implies $\varepsilon_2 \approx \beta$ when the magnitude of deformation is larger than $\sigma$, which is the case in U-U collisions.

In Fig.~\ref{fig:4} I present as well the results of \trento{} simulations of other systems, whose nuclear density parameters are listed in Tab.~\ref{tab:1}. The enhancement of elliptic flow at low $\bar p_t$ is observed, as expected, in collisions of prolate $^{129}$Xe nuclei ($\kappa_2=0.23$~\cite{Giacalone:2017dud}), recently collided at the LHC. In collisions of spherical $^{208}$Pb nuclei ($\kappa_2=0.24$~\cite{Giacalone:2017dud}), $v_2$ is instead essentially flat, with only a slight increase of $v_2$ with $\bar p_t$ due to the increasing impact parameter. The same trend is observed in collisions of mildly-oblate $^{197}$Au nuclei ($\kappa_2=0.16$~\cite{Giacalone:2018apa}) in panel (f). Note that for oblate nuclei ($\beta<0$), the enhancement of elliptic flow should occur in the large-$\bar p_t$ tail. However, with $\beta=-0.13$ the effect is not visible, as it is smeared by the sizable impact parameter.
\begin{table}[b]
\begin{tabular}{|c|c|c|c|}
\hline
species& $a$ [fm] & $R$ [fm] & $\beta$ \cr
\hline
$^{238}$U~\cite{Shou:2014eya} & 0.60 & 6.80 & 0.30\cr
$^{208}$Pb~\cite{DeJager:1987qc} & 0.55 & 6.62 & 0 \cr
$^{197}$Au~\cite{DeJager:1987qc} & 0.53 & 6.40 & -0.13~\cite{Moller:2015fba} \cr
$^{129}$Xe~\cite{Acharya:2018hhy} & 0.59 & 5.40 & 0.18~\cite{Acharya:2018hhy}  \cr
\hline
\end{tabular}
\caption{\label{tab:1} 
Parameters used in Eq.~(\ref{eq:density}) for different species.
}
\end{table}

Note however that the results presented in Fig.~\ref{fig:4} are meant to provide a qualitative description of future experimental data, and should not be intended as quantitative predictions, for two main reasons. First, they assume that $\bar p_t$ and $R$ are in a one-to-one correspondence. In hydrodynamics this is a good approximation~\cite{Bozek:2017elk}, but this relation is smeared by entropy density fluctuations in the initial state, which will eventually reduce the correlations observed in Fig.~\ref{fig:4}. Analogously, the measurement proposed in this paper requires the evaluation of $\bar p_t$ on an event-by-event basis. Since experimentally one only observes a number of particles of order $N\approx 10^3$, the determination of $\bar p_t$ in a single event is affected by statistical fluctuations which are naturally of order $1/\sqrt{N}$. These fluctuations is essentially as large as the dynamical fluctuation of $\bar p_t$ studied here (e.g. in Fig.~\ref{fig:3}, top panel), and they will yield an additional flattening of the correlations plotted in Fig.~\ref{fig:4}. The inclusion of these effects will, therefore, be crucial for a quantitative description of future experimental data.

\section{Conclusion}

In summary, the selection of events based on $\bar p_t$ allows to isolate fully-overlapping body-body collisions. The elliptic flow of the hadrons emitted from such configurations carries information about the deformation of the colliding species, thus paving the way for a new phenomenology of nuclear structure at particle colliders.  If confirmed by experiments, the present method will indeed allow to perform new experimental tests of ab-initio calculations of nuclear structure. Combined in particular with the great versatility of RHIC, it will permit to reveal the quadrupole deformation of a potentially very large number of stable nuclides. 
 
 \section{Acknowledgments}
I acknowledge the kind hospitality of the physics department of the Brookhaven National Laboratory where most of this manuscript was written. I thank  Fernando Gardim, Shengli Huang, Jean-Yves Ollitrault, Vittorio Som\`a, and Prithwish Tribedy for useful discussions.

\end{document}